# Stronger femtosecond excitation causes slower electron-phonon coupling in silicon

A. B. Swain[1], J. Kuttruff[1], J. Vorberger[2], P. Baum[1]

[1] *Universität Konstanz, Konstanz, Germany*

[2] *Helmholtz-Zentrum Dresden-Rossendorf, Dresden, Germany*

**Abstract: Electron-hole pairs in semiconductors are essential for solar cells and fast electronic circuitry, but the competition between carrier transport and relaxation into heat limits the efficiency and speed. Here we use ultrafast electron diffraction with terahertz pulse compression to measure the electron-phonon decay rate in single-crystal silicon as a function of laser excitation strength. We find that the excited electrons relax slower into phonons for higher carrier densities. The electron-phonon scattering rate changes in a nonlinear way from 400 fs at $\sim 2 \times 10^{20}/cm^3$ to 1.2 ps at $\sim 4 \times 10^{20}/cm^3$. These results indicate that a hot electron gas quenches the scattering into phonons in a temperature-dependent way. Ultrafast electronic circuitry of silicon therefore should work faster and provide higher bandwidths at lower carrier densities.**

Semiconductor materials are fundamental components of modern electronic circuits and play a critical role in various applications, such as solar cells [1], sensors [2], and nanophotonic light emission technologies [3, 4]. In these devices, electron-hole pairs exhibit complex and nonlinear motion between electrodes to achieve the desired functionality. High carrier mobility and minimal recombination processes are thus advantageous for enhancing device performance. Ballistic transistors [5] and optoelectronic devices designed for ultrafast photo-detection [6-10] are of particular interest due to their ability to allow electrons and holes to propagate with minimal scattering. In these systems, charge carriers travel across the device with nearly ballistic motion, meaning they experience very few collisions or scattering events, which enhances their speed and efficiency. Such characteristics are crucial for developing high-performance electronics and photonic devices operating at ultrafast speeds.

Generally, one of the most detrimental processes is electron-phonon coupling [11-13] in which electrons at elevated energies in the conduction band scatter and relax into phonons that ultimately convert to heat. Investigating the fundamental physics behind this phenomenon is therefore relevant for modern device design.

So far, researchers have found in many materials an electron-phonon scattering rate that is almost independent of the excited electron-hole pair density. For example, germanium (Ge) has an electron-phonon scattering rate of ~1 ps [14], while bismuth (Bi) exhibits an isotropic electron-phonon scattering rate of ~12 ps [8] which remains constant regardless of excitation strength. The two-dimensional material gadolinium telluride ($GdTe_3$) has an electron-phonon scattering rate of ~5 ps [15]. Graphene has a scattering rate with no clear relation to excitation strength [16]. In contrast, relaxation in indium antimonide (InSb) becomes quicker with increasing excitation density [17]. Likewise, bismuth (Bi) and silicon (Si) close to non-thermal melting conditions becomes quicker with increased excitation density [18, 19]. However, copper (Cu) shows a small slowdown of the time constant with increasing excitation density [20]. In some materials, the electron-phonon coupling rate also depends on the crystal plane, for example in molybdenum diselenide ($MoSe_2$) [21].

Silicon is one of the most widely used materials in solar cells and ultrafast electronic circuits [22-24]. However, the relationship between its electron-phonon scattering rate and the carrier density has not been thoroughly explored, because it requires access to crystal disorder on femtosecond scales. Here we use ultrafast single-electron diffraction [25] with terahertz-compressed electron beams [26] to probe this atomic dynamics. Figure 1a shows the experiment. The material under investigation is an ultrathin membrane of single-crystalline silicon in (100) orientation (Silson) with 30 nm thickness. Electron-hole pairs are generated by an Yb:YAG laser system (Pharos, Light Conversion UAB) that is frequency-tripled to a wavelength of 343 nm or a photon energy of 3.6 eV. The pump laser beam radius on the silicon sample is ~160 μm, the pulse duration is ~200 fs and the repetition rate is 40 kHz. Under these conditions, the sample cools back between successive excitation events [27]. Femtosecond electron pulses for ultrafast electron diffraction (UED) are generated by two-photon photoemission from a gold cathode and subsequent acceleration to a kinetic energy of 70 keV. We use ~3 electrons per pulse to minimize space charge effects [28]. For a better temporal resolution, the electron pulses are compressed in time by terahertz radiation [26] at an ultrathin aluminum membrane [29]. The probe electron beam diameter is 70 μm. Electron diffraction is probed via Bragg diffraction as a function of delay time and excitation strength.

We use 60 time steps and scan 75 times over the delay range. The integration time is ~6 s per diffraction image. Bragg spot intensity is determined by integration of a region three times larger than the spot width. Normalization to the measured Bragg spot intensity before laser incidence provides the relative change.

Figure 1b shows a terahertz streaking characterization of the compressed electron pulses at the specimen. We infer an electron pulse duration of 60 fs. Figure 1c depicts the static electron diffraction pattern of our silicon membrane. We see {220} and {400} Bragg spots; lower-order spots are symmetry-forbidden and higher-order spots are outside of the camera area. Figure 1d shows the rocking curve of the (220) Bragg spot, that is, the measured Brag spot intensity as a function of angle deviation, $\alpha$. We see the expected $\text{sinc}^2$ function with no additional interferences. Multiple scattering is therefore not relevant to our experiments [30].

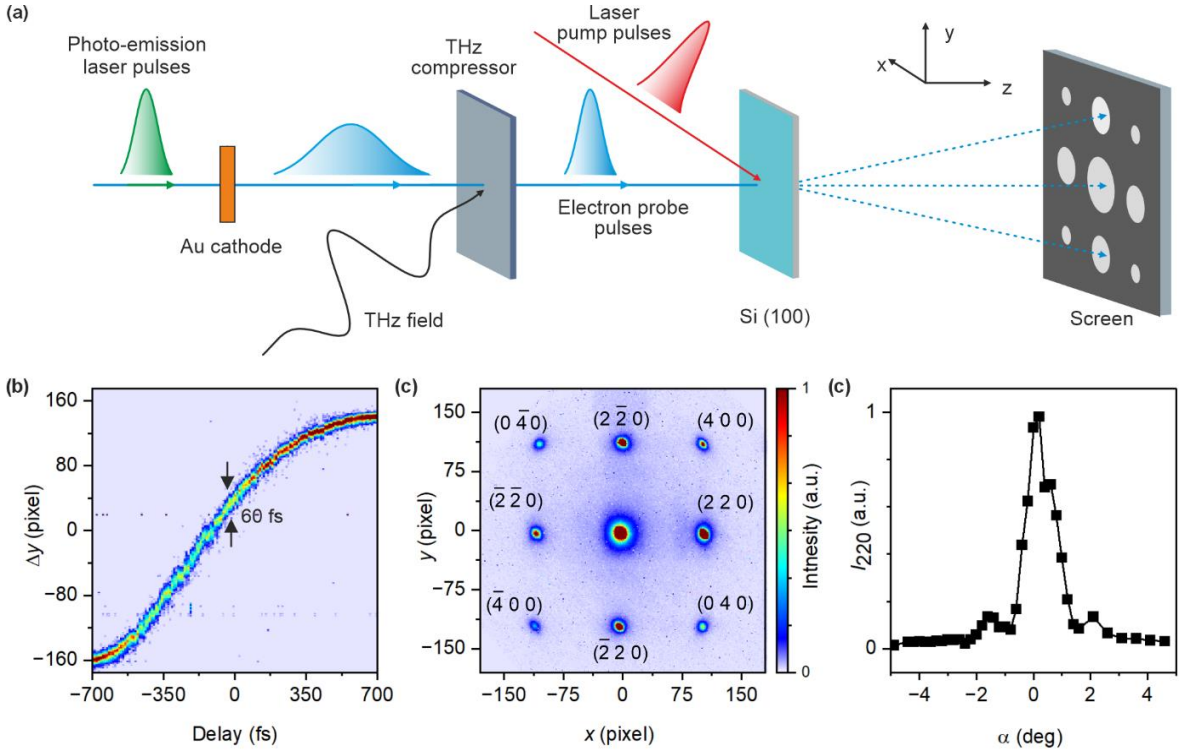

**Fig. 1:** *(a) Experimental layout of the ultrafast electron diffraction experiments with terahertz electron compression: A laser source synchronously drives electron pulse generation (green) and also produces terahertz pulses for electron compression (black). Electron pulses at an energy of 70 keV are focused using magnetic lens and compressed down to a pulse duration of 60 fs using the terahertz compressor. For the pump-probe electron diffraction measurement, a Si(001) membrane is pumped by a laser pulse*

*with a wavelength of 343 nm. (b) Terahertz streaking of the compressed electron pulse reveals the generated electron pulse duration is 60 fs. (c) The static electron diffraction of the Si(001) membrane up to the second order Bragg spots, and (d) Rocking curve of the Si(220) spot.*

Using the pump laser pulses, we create non-equilibrium electron-hole pairs and then record the arrival of energy in the lattice modes via the mean square displacement $\langle u^2 \rangle$ of the atoms around their equilibrium position. The decrease in the intensity of a Bragg spot is described by the Debye-Waller factor $\frac{I}{I_0} = exp\left[\frac{-1}{3} G^2 \cdot \langle u^2 \rangle\right]$, where $I$ is the Bragg spot intensity after laser excitation, $I_0$ is the intensity before laser arrives and $G$ is the reciprocal lattice vector of the material. When we record time-dependent Debye-Waller effects for varying laser fluence, we see the dynamics of the atomic disorder as a function of the carrier density.

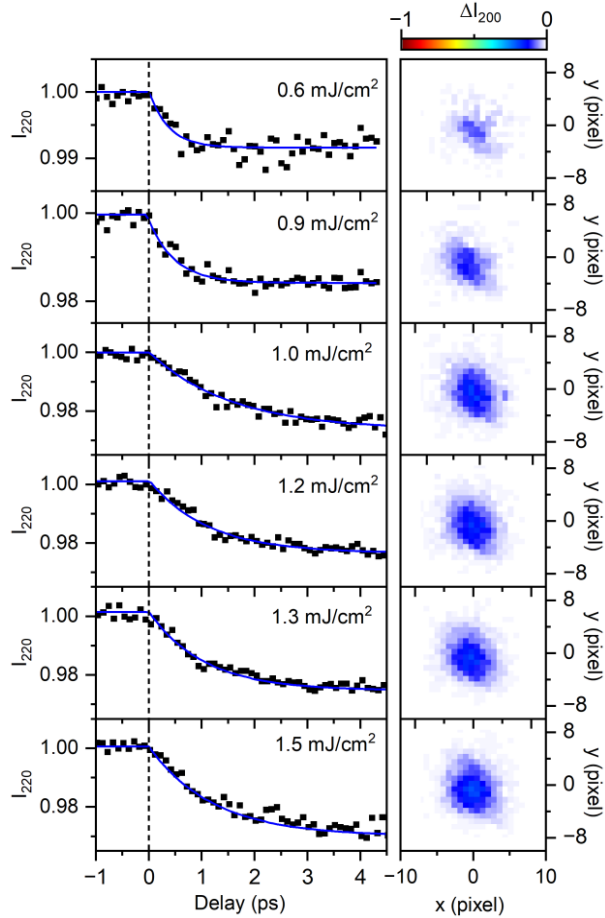

**Fig. 2:** *Ultrafast lattice dynamics in silicon. Left column, measured Bragg spot intensity changes $I_{220}$ as a function of delay in time as a function of pump fluence between 0.6-1.5 mJ/cm². The solid blue lines are the fit results. Right column, changes of the*

*diffraction pattern around the Si(220) spot. Averaged images before time zero are subtracted from averaged images between 2-4 ps after laser excitation.*

Figure 2 shows in the left panel the intensity changes of the (220) Bragg spot for various laser excitation strengths between 0.6-1.5 mJ/cm$^2$. These laser fluence values correspond to electron-hole pair densities of $1\times10^{20}$/cm$^3$ to $4\times10^{20}$/cm$^3$. The diffraction data (black squares) shows a rapid decrease as a function of time. The time constant is femtosecond to picoseconds, consistent with earlier measurements at constant laser excitation density [30, 31]. The right panel shows difference images of the (220) Bragg spot at a delay of 2-4 ps as compared to before time zero. We see no movements or increases of the width, only intensity changes. The silicon membrane therefore does not expand or contract, and also does not form inhomogeneous domains.

To quantify the absolute temperature increase and the electron-phonon coupling rate, we fit the experimental data with an exponential function according to $I_{220}(t) = \Delta I(e^{-(t-t_0)/\tau} - 1)H(t - t_0) + 1$, where $\tau$ is the rate constant, $H$ is the Heaviside function, $t_0$ is time zero and $\Delta I$ is the amplitude of the intensity change. The blue lines in Fig. 2 show these fit results. We first analyse the lattice temperature changes and disorder of the atoms. The dots in Fig. 3a depict the fitted long-term intensity changes $\Delta I_{220}$ as a function of the applied laser fluence $F$. We see that $\Delta I_{220}$ grows almost linearly with pump laser fluence. The maximum intensity drop is 3 % for 1.2 mJ/cm$^2$. The dashed line is our theoretical prediction for comparison. We first we calculate the deposited laser energy density $E_{pump}$ from the peak fluence at the center of the optical beam by taking into account the Gaussian beam profile, the measured reflectivity (~47%) and transmissivity (~4%) of the silicon membrane, the thickness of silicon membrane (~30 nm), the steady-state heat capacity (~0.7 kJ/kg/K) and the density of silicon (~2.3 g/cm$^3$). We obtain $E_{pump} \approx 250$ J/cm$^3$ or ~30 meV per Si atom for an incoming laser fluence of 1.5 mJ/cm$^2$. The resulting long-term temperature increase is ~150 K. From theoretical data [32, 33], we determine the slope of the Debye-Waller B-factor as a function of lattice temperature (~$0.00136\frac{Å^2}{K}$). Subsequently, we calculate the atomic displacements $\langle u^2 \rangle = \frac{3B}{8\pi^2}$ as a function of temperature. We then estimate the expected change of the Debye-Waller factor in the experiment as a function of the deposited laser density. The experiment starts at room temperature and we therefore plot the change of atomic displacement $\Delta\langle u^2 \rangle$ with respect to the initial static disorder at room temperature (~150 pm$^2$) which is not recorded in the experiment. The blue dashed line in Fig. 3a represents these theoretical results in comparison to the experiment (dots). The agreement shows that the total energy density $E_{pump}$ that is initially deposited into the electron gas equilibrates fully with the lattice within less than 5 ps.

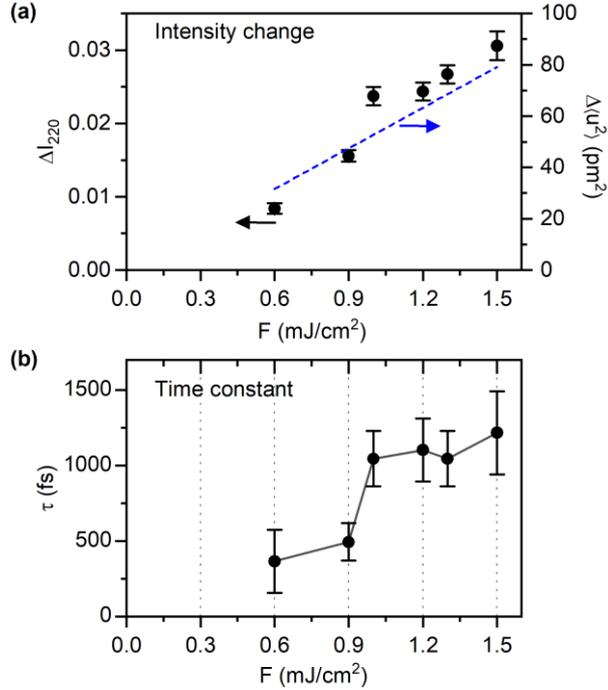

**Fig. 3:** *(a) Measured long-time change of intensity $I_{220}$ at >5 ps (black dots) and corresponding change of atomic displacements $\Delta\langle u^2\rangle$ as a function of pump laser fluence in comparison to theory (blue dashed line). (b) Measured electron-phonon time constants (black dots) as a function of pump laser fluence.*

We next analyze the electron-phonon coupling times. Figure 3b shows the fitted time constants of the experiment. The error bars are standard error of the mean. We see that all measured time constants are in the picosecond range. However, we see a faster time constant of ~400 fs for low excitation densities of less than ~1 mJ/cm$^2$ and a slower dynamic of ~1.2 ps at above ~1 mJ/cm$^2$. These results indicate that electron-phonon coupling is a not a direct rate process but depends on the absolute temperature of the electron gas.

When ultraviolet light with a photon energy above the band gap interacts with silicon, we first create a non-equilibrium distribution of high-energy electrons. In our experiment, the maximum electron energy is 3.6 eV minus band gap energy. These electrons quickly thermalize by electron-electron scattering into a hot electron gas [12, 34]. However, at the same time, energy is also transferred from the non-equilibrium electrons and from the hot electron gas into the different phonon modes of the lattice by electron-phonon scattering. Simple rate constants and two-temperature or three-temperature models [12, 34] are often insufficient to describe electron-phonon scattering in complex materials, because every part of the widely excited electronic band structure has an individual probability to scatter with all available states on the phonon dispersion relation, as far as the conservation of energy and momentum allows [12, 13].

Interestingly, in our results, we see a loss of scattering rate with elevated temperature. In non-thermal melting, the deposited laser energy is so high that it leads to the breaking of interatomic bonds into almost a liquid state [35]. Silicon at very high laser fluences, more than 50 times higher than ours, indeed shows faster dynamics at higher fluence due to changes of the potential-energy landscape of the lattice [18]. However, our measured atomic displacements $\langle u^2 \rangle$ are only 2-5% of the lattice constant of silicon, far below the Lindemann criterion [36]. Also, our Bragg diffraction patterns maintain their integrity for all applied laser fluences. Therefore, non-thermal melting does not occur. Auger recombination processes become more likely at increased carrier densities and could lead to an increase of scattering rate, but our measurement show the opposite sign. Shockley-Read-Hall recombination through defect states takes nanoseconds [37] and is therefore also not relevant for our results.

We see two possibilities to explain our results. First, the electron gas can have a nonlinear and temperature-dependent heat capacity that increases at higher temperatures. Higher laser fluences therefore create a quenched temperature increase that in turn slows down the electron-phonon coupling dynamics. Second, it is possible that the electron-phonon scattering rate depends on the details of the involved electronic states via an Eliashberg function with different phonon branches [12, 13]. Surprising remains the sign of the measured effect, that is, a slow-down of the rate at higher excitation strength.

In summary, we have investigated the electron-phonon coupling dynamics in silicon using ultrafast electron diffraction with terahertz pulse compression. The electron-phonon scattering rate decreases nonlinearly with increasing carrier density. This phenomenon suggests a temperature-dependent quenching effect that slows down the relaxation into phonons, contrary to trends observed in most other materials [14-19]. This finding may have practical implications, for example, for optimizing silicon-based photoconductive switches or optoelectronic detectors [6, 7]. Performance and bandwidth can be improved at well-chosen excitation strengths. Also, metamaterial-based solar cells use light-harvesting nanostructures to increase collection efficiency [38] and naturally produce a high local carrier density. According to our results, optimized carrier densities are beneficial in terms of lifetime. The results also highlight the utility of ultrafast electron diffraction with terahertz compression as a useful tool to probe the temperature-dependent dynamics of atomic disorder in crystalline materials.

**Acknowledgement**: We acknowledge financial support by the Deutsche Forschungsgemeinschaft via SFB 1432 and the European research executive agency (REA) through Horizon-MSCA-2021-PF-01 proposal 101064961. We thank Eruthuparna Ramachandran for assistance in the laboratory.

**References**

[1] C. Ballif, F.-J. Haug, M. Boccard, P. J. Verlinden, G. Hahn, Status and perspectives of crystalline silicon photovoltaics in research and industry. *Nat. Rev. Mater.* **7**, 597–616 (2022).

[2] T. Dutta, T. Noushin, S. Tabassum, S. K. Mishra, Road map of semiconductor metal-oxide-based sensors: A review. *Sensors* **23**, 6849 (2023).

[3] A. Mischok, S. Hillebrandt, S. Kwon, M. C. Gather, Highly efficient polaritonic light-emitting diodes with angle-independent narrowband emission. *Nat. Photonics* **17**, 393–400 (2023).

[4] M. W. Swift, A. L. Efros, S. C. Erwin, Controlling light emission from semiconductor nanoplatelets using surface chemistry. *Nat. Commun.* **15** (2024).

[5] X. Du, I. Skachko, A. Barker, E. Y. Andrei, Approaching ballistic transport in suspended graphene. *Nat. Nanotechnol.* **3**, 491–495 (2008).

[6] Z. H. Li, J. X. He, X. H. Lv, L. F. Chi, K. O. Egbo, M.-D. Li, T. Tanaka, Q. X. Guo, K. M. Yu, C. P. Liu, Optoelectronic properties and ultrafast carrier dynamics of copper iodide thin films. *Nat. Commun.* **13** (2022).

[7] N. P. Gallop, D. R. Maslennikov, N. Mondal, K. P. Goetz, Z. Dai, A. M. Schankler, W. Sung, S. Nihonyanagi, T. Tahara, M. I. Bodnarchuk, M. V. Kovalenko, Y. Vaynzof, A. M. Rappe, A. A. Bakulin, Ultrafast vibrational control of organohalide perovskite optoelectronic devices using vibrationally promoted electronic resonance. *Nat. Mater.* **23**, 88–94 (2023).

[8] V. Tinnemann, C. Streubühr, B. Hafke, A. Kalus, A. Hanisch-Blicharski, M. Ligges, P. Zhou, D. Von Der Linde, U. Bovensiepen, M. Horn-Von Hoegen, Ultrafast electron diffraction from a Bi(111) surface: Impulsive lattice excitation and Debye-Waller analysis at large momentum transfer. *Struct. Dyn.* **6**, 1–9 (2019).

[9] D. Lin, S. Li, J. Wen, H. Berger, L. Forró, H. Zhou, S. Jia, T. Taniguchi, K. Watanabe, X. Xi, M. S. Bahramy, Patterns and driving forces of dimensionality-dependent charge density waves in 2H-type transition metal dichalcogenides. *Nat. Commun.* **11**, 1–9 (2020).

[10] S. Butscher, F. Milde, M. Hirtschulz, E. Malić, A. Knorr, Hot electron relaxation and phonon dynamics in graphene. *Appl. Phys. Letts.* **91** (2007).

[11] A. Ruckhofer, D. Campi, M. Bremholm, P. Hofmann, G. Benedek, M. Bernasconi, W. E. Ernst, A. Tamtögl, Terahertz surface modes and electron-phonon coupling on $Bi_2Se_3$. *Phys. Rev. Res.* **2** (2020).

[12] F. Caruso, D. Novko, Ultrafast dynamics of electrons and phonons: from the two-temperature model to the time-dependent Boltzmann equation. *Adv. Phys.: X* **7** (2022).

[13] L. Waldecker, R. Bertoni, R. Ernstorfer, J. Vorberger, Electron-phonon coupling and energy flow in a simple metal beyond the two-temperature approximation. *Phys. Rev. X* **6**, 1–11 (2016).

[14] K. Sokolowski-Tinten, U. Shymanovich, M. Nicoul, J. Blums, A. Tarasevitch, M. Horn-von Hoegen, D. von der Linde, A. Morak, T. Wietler, Energy relaxation and anomalies in the thermo-acoustic response of femtosecond laser-excited germanium. *Optics InfoBase Conference Papers* (2006).

[15] I. Gonzalez-Vallejo, V. L. Jacques, D. Boschetto, G. Rizza, A. Hadj-Azzem, J. Faure, D. Le Bolloc'H, Time-resolved structural dynamics of the out-of-equilibrium charge density wave phase transition in $GdTe_3$. *Struct. Dyn.* **9** (2022).

[16] C. Lee, A. Marx, G. H. Kassier, R. J. Miller, Disentangling surface atomic motions from surface field effects in ultrafast low-energy electron diffraction. *Commun. Mater.* **3**, 1–9 (2022).

[17] P. B. Hillyard, K. J. Gaffney, A. M. Lindenberg, S. Engemann, R. A. Akre, J. Arthur, C. Blome, P. H. Bucksbaum, A. L. Cavalieri, A. Deb, R. W. Falcone, D. M. Fritz, P. H. Fuoss, J. Hajdu, P. Krejcik, J. Larsson, S. H. Lee, D. A. Meyer, A. J. Nelson, R. Pahl, D. A. Reis, J. Rudati, D. P. Siddons, K. Sokolowski-Tinten, D. von der Linde, J. B. Hastings, Carrier-density-dependent lattice stability in insb. *Phys. Rev. Lett.* **98** (2007).

[18] M. Harb, R. Ernstorfer, C. T. Hebeisen, G. Sciaini, W. Peng, T. Dartigalongue, M. A. Eriksson, M. G. Lagally, S. G. Kruglik, R. J. D. Miller, Electronically driven structure changes of Si captured by femtosecond electron diffraction. *Phys. Rev. Lett.* **100** (2008).

[19] G. Sciaini, M. Harb, S. G. Kruglik, T. Payer, C. T. Hebeisen, F.-J. M. z. Heringdorf, M. Yamaguchi, M. H.-v. Hoegen, R. Ernstorfer, R. J. D. Miller, Electronic acceleration of atomic motions and disordering in bismuth. *Nature* **458**, 56–59 (2009).

[20] M. Z. Mo, V. Becker, B. K. Ofori-Okai, X. Shen, Z. Chen, B. Witte, R. Redmer, R. K. Li, M. Dunning, S. P. Weathersby, X. J. Wang, S. H. Glenzer, Determination of the electron-lattice coupling strength of copper with ultrafast MeV electron diffraction. *Rev. Sci. Instrum.* **89** (2018).

[21] M. F. Lin, V. Kochat, A. Krishnamoorthy, L. Bassman, C. Weninger, Q. Zheng, X. Zhang, A. Apte, C. S. Tiwary, X. Shen, R. Li, R. Kalia, P. Ajayan, A. Nakano, P. Vashishta, F. Shimojo, X. Wang, D. M. Fritz, U. Bergmann, Ultrafast non-radiative dynamics of atomically thin $MoSe_2$. *Nat. Commun.* **8**, 1–7 (2017).

[22] M. Sui, Y. Chu, R. Zhang, A review of technologies for high efficiency silicon solar cells. *J. Phys. Conf. Ser.* **1907**, 012026 (2021).

[23] R. Saive, Light trapping in thin silicon solar cells: A review on fundamentals and technologies. *Prog. Photovoltaics Res. Appl.* **29**, 1125–1137 (2021).

[24] M. A. Foster, R. Salem, D. F. Geraghty, A. C. Turner-Foster, M. Lipson, A. L. Gaeta, Silicon-chip-based ultrafast optical oscilloscope. *Nature* **456**, 81–84 (2008).

[25] P. Baum, On the physics of ultrashort single-electron pulses for time-resolved microscopy and diffraction. *Chem. Phys.* **423**, 55–61 (2013).

[26] C. Kealhofer, W. Schneider, D. Ehberger, A. Ryabov, F. Krausz, P. Baum, All-optical control and metrology of electron pulses. *Science* **352**, 429–433 (2016).

[27] D. Kazenwadel, N. Neathery, S. Prakash, A. Ariando, P. Baum, Cooling times in femtosecond pump-probe experiments of phase transitions with latent heat. *Phys. Rev. Res.* **5** (2023).

[28] M. Aidelsburger, F. O. Kirchner, F. Krausz, P. Baum, Single-electron pulses for ultrafast diffraction. *Proceedings of the National Academy of Sciences* **107**, 19714–19719 (2010).

[29] D. Ehberger, K. J. Mohler, T. Vasileiadis, R. Ernstorfer, L. Waldecker, P. Baum, Terahertz compression of electron pulses at a planar mirror membrane. *Phys. Rev. Appl.* **11** (2019).

[30] I. González Vallejo, G. Gallé, B. Arnaud, S. A. Scott, M. G. Lagally, D. Boschetto, P.-E. Coulon, G. Rizza, F. Houdellier, D. Le Bolloc'h, J. Faure, Observation of large multiple scattering effects in ultrafast electron diffraction on monocrystalline silicon. *Phys. Rev. B.* **97** (2018).

[31] M. Harb, W. Peng, G. Sciaini, C. T. Hebeisen, R. Ernstorfer, M. A. Eriksson, M. G. Lagally, S. G. Kruglik, R. J. D. Miller, Excitation of longitudinal and transverse coherent acoustic phonons in nanometer free-standing films of (001) Si. *Phys. Rev. B.* **79** (2009).

[32] J. S. Reid, J. D. Pirie, Dynamic deformation and the Debye-Waller factors for silicon-like crystals. *Acta Crystallogr.* **36**, 957–965 (1980).

[33] C. Flensburg, R. F. Stewart, Lattice dynamical Debye-Waller factor for silicon. *Phys. Rev. B* **60**, 284–291 (1999).

[34] Z. Lin, L. V. Zhigilei, V. Celli, Electron-phonon coupling and electron heat capacity of metals under conditions of strong electron-phonon nonequilibrium. *Phys. Rev. B* **77**, 1–17 (2008).

[35] B. J. Siwick, J. R. Dwyer, R. E. Jordan, R. J. D. Miller, An atomic-level view of melting using femtosecond electron diffraction. *Science* **302**, 1382–1385 (2003).

[36] K. Sokolowski-Tinten, C. Blome, J. Blums, A. Cavalleri, C. Dietrich, A. Tarasevitch, I. Uschmann, E. Förster, M. Kammler, M. Horn-von Hoegen, D. von der Linde, Femtosecond x-ray measurement of coherent lattice vibrations near the Lindemann stability limit. *Nature* **422**, 287–289 (2003).

[37] P. T. Webster, R. A. Carrasco, A. T. Newell, J. V. Logan, P. C. Grant, D. Maestas, C. P. Morath, Utility of Shockley–Read–Hall analysis to extract defect properties from semiconductor minority carrier lifetime data. *J. Appl. Phys.* **133** (2023).

[38] H. A. Atwater, A. Polman, Plasmonics for improved photovoltaic devices. *Nat. Mater.* **9**, 205–213 (2010).